\documentstyle[aps,epsfig,multicol]{revtex}
\draft  
\begin{document}

\title{Strange quark polarization of the nucleon: A parameter-independent
	prediction of the chiral potential model}
\author{X.B. Chen$^{a,b}$, X.S. Chen$^{a,c}$, Amand Faessler$^c$, 
	Th. Gutsche$^c$, F. Wang$^a$}
\address{$^a$Department of Physics and Center for Theoretical Physics, 
        	Nanjing University, Nanjing 210093, China\\
        $^b$Department of Physics, Changsha Institute of Electricity,
		Changsha 410077, China\\
	$^c$Institut f\"ur Theoretishe Physik, Universit\"at T\"ubingen,
	        Auf der Morgenstelle 14, D-72076 T\"ubingen, Germany}
\date{\today}
\maketitle

\begin{abstract}
We perform a one-loop calculation of the strange quark polarization 
($\Delta s$) of the nucleon in a SU(3) chiral potential model.
We find that if the intermediate excited quark states
are summed over in a proper way, i.e., summed up to a given energy instead 
of given radial and orbital quantum numbers, $\Delta s$ turns out to be almost
independent of {\em all} the model parameters: quark masses, scalar- and 
vector-potential strengths. 
The contribution from the
quark-antiquark pair creation and annihilation ``$Z$'' diagrams is found to be
significant. Our numerical results agree quite reasonably with experiments and 
lattice QCD calculations.

\end{abstract}

\begin{multicols}{2}
The intrinsic strangeness content of the nucleon is a key ingredient to 
understand the 
structure and dynamics inside baryons. While the experimental investigation of
the nucleon spin
structure \cite {spin} clearly indicates that a strange quark sea exists and 
is also polarized relative to the nucleon spin, 
the successes of the naive spin-flavor 
SU(6) valence quark model in 
various aspects suggest that the strangeness content should belong to higher 
order effects for the nucleon. The SU(3) flavor chiral quark model, which 
couples light quarks to octet pseudoscalar 
mesons by the requirement of chiral symmetry, provides a natural mechanism for 
such a perturbative picture: at zeroth order the ground state octet baryons 
are described by a SU(6) wave function of three valence quarks, 
and at second order in the quark-meson coupling 
sea quarks can be generated by emitting a meson from the valence quark. 
For example, in the nucleon the strange quark can be generated by emitting a 
$K^+$ from the $u$ quark or by emitting a $K^0$ from the $d$ quark (Fig. 1), 
and hence strange quarks can contribute to the nucleon spin structure. 

\begin{center}
\begin{minipage}{85mm}
\begin{figure}
\begin{center}
\psfig{figure=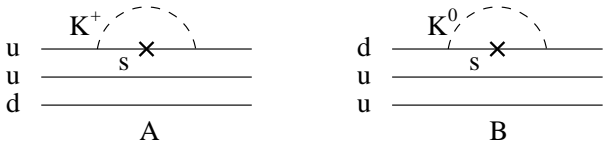,width=8cm}
\bigskip
\begin{tighten}
\caption{Strange quark contribution to the nucleon spin; 
a cross on the quark line
 denotes the axial 
vector vertex $\gamma ^3 \gamma ^5$.} 
\end{tighten}
\end{center}
\end{figure}
\end{minipage}
\end{center}

In this paper we will adopt the standard perturbation theory to calculate the
strange quark polarization of the nucleon in the framework of a SU(3) 
chiral potential model. As will be shown, up to second order
the diagrams of Fig. 1 are the only contributions. Hence, a quantitative
determination of the strange quark polarization is a clean test for the 
interaction picture of the 
chiral quark model, which allows in this case only admixtures of $K$ mesons.
On the other hand the $u$ or $d$ quark polarizations obtain 
contributions from many other diagrams and therefore depend on many more model 
parameters.

To set up our calculation scheme, we first define the effective Lagrangian 
for the SU(3) chiral potential model:
\begin{eqnarray}
{\cal L}&=&\bar{\psi} [ i\partial \hspace*{-2mm}/
		- S(r)- \gamma^0 V(r)]\psi - \nonumber \\
         &&\frac{1}{2F_\pi} \bar{\psi} [ 
	S(r)(\sigma +i\gamma ^5 \lambda^i \phi_i) + 
	(\sigma +i\gamma ^5\lambda^i \phi_i )S(r) ]\psi + \nonumber \\
         &&\frac 12 (\partial_\mu \sigma)^2 + 
           \frac 12 (\partial_\mu \phi_i)^2 -
	   \frac 12 m_\sigma^2 \sigma^2 -
	   \frac 12 m_i^2\phi_i^2.
	    \label{Lagrangian} 
\end{eqnarray}
The model Lagrangian  is derived from the $\sigma$ model in which meson fields 
are introduced to restore chiral symmetry \cite{Thomas}. 
$\psi$ is the quark field with flavor and color indices implied; 
the scalar term
$S(r)=cr+m$ represents the the linear scalar confinement potential $cr$ and
the quark
mass matrix $m$;
$V(r)=-\alpha /r$ is the Coulomb type vector potential and
$F_\pi$=93MeV is the pion decay 
constant. $\sigma$ and $\phi_i$ ($i$ runs from $1$ to $8$) are the scalar
and pseudoscalar meson fields, respectively and $\lambda_i$ are the Gell-Mann 
matrices. The quark-meson interaction term of Eq.(\ref{Lagrangian}) is
symmetrized since the
mass matrix $m$ does not commute with all $\lambda_i$
for different quark masses.

The zeroth order quark Hamiltonian is set up as 
\begin{equation}
H_q = \int d^3 x \psi ^\dagger  [\vec \alpha \cdot \frac 1i\vec \partial + 
	\beta S(r) +V(r) ] \psi.
\end{equation}
It has discrete eigenstates which are obtained by
numerical solution of the Dirac equation with a scalar and vector field
\cite{Gutsche}. We write the solution as: 
\begin{equation}
\psi(x) =\sum _\alpha u_\alpha(x)a_\alpha +
        \sum _\beta v_\beta (x)b^\dagger _\beta.
 	 \label{basis}
\end{equation}

Eq. (\ref{basis}) forms the basis of our unperturbed wave functions, where
quarks are bound permanently by the confinement potential 
which is included in $H_q$.  
From Eq. (\ref{basis}) we can construct the quark propagator:  
\begin{eqnarray}
D(x_1,x_2)&\equiv& \langle 0|T\{\psi(x_1),\bar\psi (x_2)\}|0\rangle \nonumber\\
        &=& \theta(t_1-t_2)\sum _\alpha u_\alpha(x_1) \bar{u_\alpha}(x_2)-
        \nonumber \\
        &&  \theta(t_2-t_1)\sum _\beta v_\beta(x_1) \bar{v_\beta}(x_2),  
\end{eqnarray}
The meson propagator given by Eq. (\ref{Lagrangian}) is the free one:
\begin{eqnarray}
\Delta_{ij}(x_1,x_2)&\equiv&\langle 0|T\{\phi_i(x_1),\phi_j(x_2)\}|0\rangle 
        \nonumber \\
        &=& \frac{i}{(2\pi)^4} \int d^4 q 
                \frac{\delta _{ij} e^{-iq\cdot(x_1-x_2)}}
                {q^2-m_i^2+i\epsilon}. 
\end{eqnarray}

Given the unperturbed basis we can construct any physical quantity up to a
desired order in the quark-meson interaction. In the following we are 
studying the quark contribution of flavor $q$ ($q=u,d,s$) to the 
nucleon spin which is defined through
\begin{equation}
\Delta q =\frac{\langle N| \int d^3 x\bar{\psi_q} \gamma^3 \gamma^5 \psi_q
		|N\rangle }
	 {\langle N | N \rangle }. \label{Deltaq}
\end{equation}

At zeroth order $H_q$ gives the usual SU(6) three-quark states for the nucleon 
with
the single quark wave function $u_\alpha$ in the ground state.
The zeroth order diagram for the numerator of Eq. (\ref{Deltaq}) is indicated in
Fig. 2A, and the denominator by the diagram of Fig. 3A, which is simply unity.
Clearly, strange quarks do not contribute at this order.

The corresponding Feynman diagrams which contribute to  $\Delta s$ up to second
order are shown in Figs. 2 and 3.
The denominator $\langle N|N\rangle$ can be denoted as
$(1+{\rm const.}/F_\pi ^2)$, which can be expanded
$(1-{\rm const.}/F_\pi ^2 + \cdots)$ and has then to be
multiplied with the numerator
$\langle N| \int d^3 x\bar\psi \gamma^3 \gamma^5 \psi|N\rangle $.
If finally, only terms of order $1/F_\pi ^2$ are kept in the product of the
normalization and the matrix element of the spin,
this has no effect on $\Delta s$, since
already the lowest order admixture of $s$ quark is proportional to
$1/F_\pi ^2$.

\begin{center}
\begin{minipage}{85mm}
\begin{figure}
\begin{center}
\psfig{figure=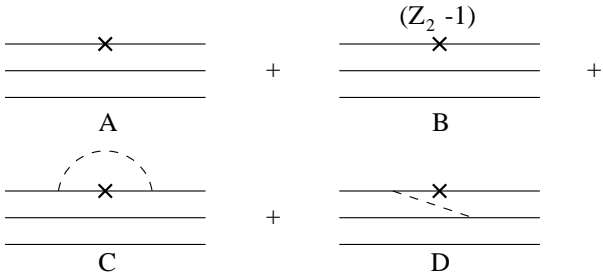,width=8cm}

\bigskip

\begin{tighten}
\caption{Feynman diagrams for the matrix element
$\langle N| \int d^3 x\bar\psi \gamma^3 \gamma^5 \psi|N\rangle$ 
up to second order;
A is of the zeroth
order, B is the renormalization counter term, C and D are vertex and exchange
diagrams respectively. The meson line in C can be a $\pi$ and $\eta$ (while the
intermediate quark is $u$ or $d$), or a $K$ (while the intermediate quark
is $s$); the meson line in D can only be a $\pi$ or $\eta$.}
\end{tighten}
\end{center}
\end{figure}
\end{minipage}
\end{center}

\begin{center}
\begin{minipage}{85mm}
\begin{figure}
\begin{center}
\psfig{figure=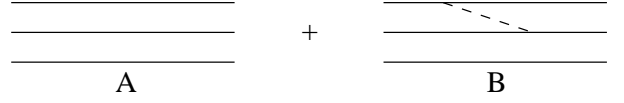,width=8cm}

\bigskip

\begin{tighten}
\caption{Feynman diagrams for the normalization
$\langle N|N\rangle$ up to second order; A is of the zeroth
order which is simply unity, B is the meson exchange
diagram. The meson line in B is a $\pi$ or $\eta$.}
\end{tighten}
\end{center}
\end{figure}
\end{minipage}
\end{center}

In the Lagrangian of Eq. (\ref{Lagrangian}) 
the main effect of the nonperturbative quark-gluon interaction 
is supposed to be included by the scalar and vector
potentials. In principle we can also include a
residual perturbative gluon piece. This will 
introduce further modifications
on $\Delta u$ and $\Delta d$. However since the perturbative quark-gluon 
interaction is diagonal in flavor space, it cannot generate 
strange quark admixtures in second order for the nucleon. 

Now we are in a good position to calculate $\Delta s$ for the nucleon: up to 
second order the only diagrams we need to consider are the subset of the
diagrams of Fig. 2, which are given in Fig. 1. For the evaluation,
we first give the explicit form
for $u_\alpha(x)$ and $v_\beta (x)$ with:  $u_\alpha(x)=
e^{-iE_\alpha t}u_\alpha (\vec x)\tau_\alpha$, 
$v_\beta (x)=e^{iE_\beta t}v_\beta (\vec x)\tau_\beta$, where $\tau$ is the 
flavor wavefunction and the spatial wavefunction is:
\begin{equation}
u_\alpha (\vec x)=
\left(
\begin{array}{c}
g_{njl} \\
-i \vec \sigma \cdot \hat{\vec r} f_{njl}
\end{array}
\right)
Y_{jl}^m =
\left(
\begin{array}{c}
g_{njl}Y_{jl}^m \\
if_{njl}Y_{jl'}^m 
\end{array}
\right), \label{u}
\end{equation}
where $g$ and $f$ are real functions, $n$ is the radial quantum number, 
and $Y_{jl}^m (\hat{\vec r})$ are the vector spherical harmonics.   
The second equality of Eq.(\ref{u}) follows from 
$\vec \sigma \cdot \hat{\vec r}Y_{jl}^m =-Y_{jl'}^m$ 
with $l'=2j-l$. For computational 
convenience, we will use exactly the same form for 
$v_\beta (x)$. Since for the antiquark solution the lower component is the 
large component, for $v_\beta (x)$ $l$ is actually the orbital quantum 
number of the 
small component, and $|E_{j=l+1/2}|>|E_{j=l-1/2}|$.  
Thus for the antiquarks the sequence is inversed.

Denoting the initial and final quark states as $u_i$ and $u_f$ 
respectively, the contribution of the diagrams of Fig. 1 is:
\begin{eqnarray}
\delta s&=&\frac{1}{F_\pi^2}
	\int d^3x d^4x_1 d^4x_2 \Delta(x_2,x_1)
	\bar u_f(x_2) S(r_2)\gamma ^5 \lambda ^i 
	\times \nonumber \\
	&& D(x_2,x) \gamma^3\gamma^5 D(x,x_1) 
 	S(r_1)\gamma^5 \lambda^i u_i(x_1). \label{Fig1}
\end{eqnarray}
Here we use $\delta s$ to indicate that it is only the 
contribution from a single quark state. 
Inserting the explicit expressions for the propagators, we get
\begin{eqnarray}
\delta s&=&\frac{1}{F_\pi^2}
	\int d^4x_1 d^4x_2 \bar u_f(x_2) S(r_2)\gamma ^5 \lambda^i 
        \times \nonumber \\
        && \left[\theta(t_2-t)\theta(t-t_1)\sum _{\alpha\alpha'}u_\alpha(x_2)
	\Delta_{\alpha\alpha'}
		\bar u_{\alpha'}(x_1)+ \right.\nonumber \\
	&&~ \theta(t_1-t)\theta(t-t_2)\sum_{\beta\beta'}v_\beta(x_2)
	\Delta_{\beta\beta'}
		\bar v_{\beta'}(x_1) - \nonumber \\
	&&~\theta(t_2-t)\theta(t_1-t)\sum_{\alpha\beta'} u_\alpha(x_2)
	\Delta_{\alpha\beta'}
                \bar v_{\beta'}(x_1) -\nonumber \\
        &&~\left.
	\theta(t-t_2)\theta(t-t_1)\sum _{\beta \alpha'}v_\beta (x_2)
	\Delta_{\beta\alpha'}
                \bar u_{\alpha'}(x_1)\right]\times  \nonumber \\
        && S(r_1)\gamma^5 \lambda^i u_i(x_1) \frac{i}{(2\pi)^4} \int d^4 q 
                \frac{\delta _{ij} e^{-iq\cdot(x_1-x_2)}}
                {q^2-m_i^2+i\epsilon}
		\label{expr1}.
\end{eqnarray}
where 
$\Delta_{\alpha\alpha'}=
	\int d^3x \bar u_\alpha \gamma^3\gamma^5 u_{\alpha'}$,
and similarly for 
$\Delta_{\beta\beta'}$ etc. 
The four time-ordered terms in Eq.(\ref{expr1}) correspond to the
time-ordered diagrams of Fig. 4.

\begin{center}
\begin{minipage}{85mm}
\begin{figure}
\begin{center}
\psfig{figure=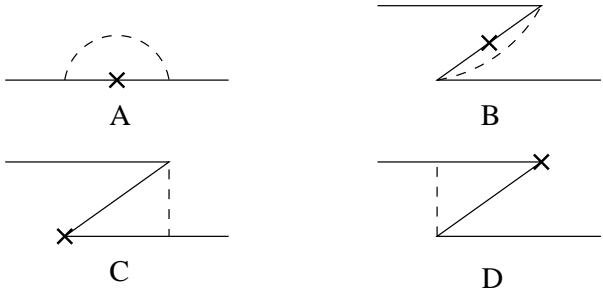,width=8cm}

\bigskip

\begin{tighten}
\caption{Time-ordered diagrams of Fig. 1; A is the positive-energy
 state contribution;
B is the negative-energy state contribution;
C and D are the quark-antiquark pair
creation and annihilation ``$Z$'' diagrams.}
\end{tighten}
\end{center}
\end{figure}
\end{minipage}
\end{center}

We omit here the details for calculating $\delta s$ of Eq. (\ref{expr1}).
The integrals of Eq. (\ref{expr1}) can be reduced analytically to
radial integrations at the vertex points ($r_1$ and $r_2$) and of loop
momentum $|\vec q|$.
The the remaining integrations are carried out
numerically. $\Delta s$ for the whole nucleon is just $\delta s$
times a 
spin-isospin factor which can be straightforwardly calculated to be 2.

In Table I we list our model parameters. 
Since $F_\pi=93$MeV and $m_K=495$Mev are fixed by experiment, 
our model contains four free parameters: 
the two quark masses $m_{u,d}$, $m_s$ and the two strength constants 
of the scalar and 
vector potential denoted by $c$ and $\alpha$. The parameter $\alpha$
is fixed by the long-wavelength, transverse fluctuations of the QCD 
based static-source flux-tube picture \cite{flux1,flux2}. It was obtained to 
be $0.26$ in \cite{alpha1} and $0.30$ in \cite{alpha2}, while a much larger
value of about $0.52$ was used by the Cornell group \cite{Cornell}.
Recent lattice calculation \cite{Bali}
got a value around $0.32$ in the 
quenched approximation, and suggested that relaxing the quenched
approximation may lead to $\alpha \sim 0.40$.
Quark masses and confinement strength are rather uncertain 
quantities. To study the variation of $\Delta s$ over all the parameters,
we choose in our calculation four different sets of parameters,  
including both current and constituent quark masses. 

We study very 
different parameters because they often vary significantly
from one model to another. For this model one possible choice of parameters 
to produce the correct 
nucleon mass, $g_A$, etc.  
is given in Ref. \cite{Chen}. 

Table II gives the numerical results of $\Delta s$ for the first two sets of
parameters.
The intermediate quark/antiquark
states are summed over up to a radial quantum number of $n=8$ and total
angular momentum $j=17/2$. We also list
the intermediate results with the summation including states up to
$n=6$ and $j=11/2$. The contributions
from the four time-ordered diagrams in Fig. 4 are given separately. We
note significant contributions from Fig. 4C and Fig. 4D, in which a
quark-antiquark pair is created or annihilated by the axial vector current;
these processes are usually referred to as the ``$Z$'' diagrams.
On the other hand the diagram of Fig. 4B gives a fairly large positive
contribution, therefore if the ``$Z$'' diagrams are neglected we
would incorrectly conclude that $\Delta s$ in the nucleon is positive.

From Table II one would conclude that a stronger confinement also gives a larger
$\Delta s$. This is due to the coupling of the meson field to the quark field
which is proportional to the effective quark mass $S(r)=cr+m$. However, to 
compare with the energy scale in the 
lattice QCD calculation of $\Delta s$, we should sum the excited states up to 
a given energy instead of given radial and orbital quantum number. The
resummed $\Delta s$ according to energy 
are given in Fig. 5. Since the quark states are discrete, we get plateaus in 
Fig. 5 at the energies where no new states emerge.

\begin{table}
\begin{center}
\begin{tighten}
\caption{Model parameters}
\end{tighten}
\begin{scriptsize}
\begin{tabular}{ccccc}
para. &$m_{u,d}$ &$m_s$  &$\alpha$ & $c$ \\
set   &[MeV]&[MeV]& &[GeV$^2$] \\ \hline
 1    &$10$ &$150$&$0.26$&$0.11$ \\
 2    &$10$ &$150$&$0.26$&$0.16$ \\
 3    &$300$&$500$&$0.26$&$0.11$ \\
 4    &$10$ &$150$&$0.50$&$0.18$ 
\end{tabular}
\end{scriptsize}
\end{center}
\end{table}

\vspace*{-6mm}
\begin{table}
\begin{center}
\begin{tighten}
\caption{Numerical results for $\Delta s$
 by summing the intermediate quark/antiquark
states up to given radial and orbital
quantum numbers.}
\end{tighten}
\begin{scriptsize}
\begin{tabular}{cccccccc}
$n$&$j$ &set&Fig. 4A &Fig. 4B &Fig. 4C &Fig. 4D & sum \\ \hline
6  &$11/2$&1&$-.0212$ &$+.0829$ &$-.0844$ &$-.0844$ &$-.1070$ \\
   &      &2&$-.0340$ &$+.1238$ &$-.1255$ &$-.1255$ &$-.1613$ \\ \hline
8  &$17/2$&1&$-.0220$ &$+.0964$ &$-.1109$ &$-.1109$ &$-.1475$ \\
   &      &2&$-.0345$ &$+.1445$ &$-.1655$ &$-.1655$ &$-.2211$
\end{tabular}
\end{scriptsize}
\end{center}
\end{table}

\vspace*{-6mm}
\begin{table}
\begin{center}
\begin{tighten}
\caption{Numerical results for $\Delta s$
by summing over the intermediate quark/antiquark
states up to the energy 1.7GeV.}
\end{tighten}
\begin{scriptsize}
\begin{tabular}{cccccc}
para. set&Fig. 4A &Fig. 4B &Fig. 4C &Fig. 4D & sum \\ \hline
1        & $-.0228$ &$+.0785$  &$-.0742$ &$-.0742$ &$-.0927$ \\
2        & $-.0365$ &$+.0958$  &$-.0805$ &$-.0805$ &$-.1017$ \\
3        & $-.0464$ &$+.1538$  &$-.1046$ &$-.1046$ &$-.1018$ \\
4        & $-.0400$ &$+.0834$  &$-.0636$ &$-.0636$ &$-.0838$
\end{tabular}
\end{scriptsize}
\end{center}
\end{table}
\clearpage

Since the strange axial current is a non-conserved composite operator, it has
divergent matrix element (as is seen in Fig. 5), and therefore must be
renormalized. Analogous to the lattice renormalization, we cut the quark 
intermediate states at an energy of $1.7$GeV, which is roughly the
inverse of the lattice spacing in the lattice calculation of 
$\Delta s$ ($a^{-1}=1.74$GeV in \cite{Latt2}). 
The ``renormalized'' results are given in Table III. 
(In principle, we can also do renormalization by imposing a cutoff
on the meson momentum, such as using the Pauli-Villars regulator  
$(q^2-\Lambda^2)^{-1}$. 
But then the quark intermediate states have to be summed 
up to convergence. 
In practice this is not workable.
An illustration with $\Lambda =1.7$GeV is given in Fig. 5.)

\begin{center}
\begin{minipage}{85mm}
\begin{figure}
\begin{center}
\psfig{figure=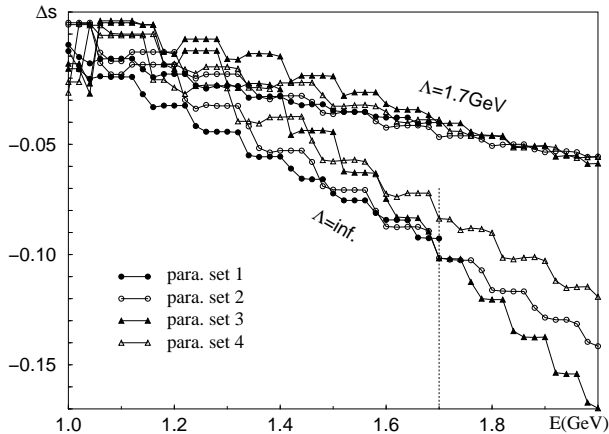,width=8cm}

\bigskip

\begin{tighten}
\caption{plot of $\Delta s$ as a function of the maximal energy up to which
the intermediate states are summed.}
\end{tighten}
\end{center}
\end{figure}
\end{minipage}
\end{center}

We note a very interesting phenomena in Fig. 5: the result for 
$\Delta s$ summed up to a given energy is rather robust against the variation 
of {\em all} the parameters. 
The insensitivity is especially impressive compared to the huge variation
of $m$ and $c$. 
Table III shows that the ``$Z$'' diagram's contribution is still
significant.

The insensitivity of $\Delta s$ on the parameter sets can 
be attributed to the fact that
the increase of $m$ and $c$ (see Eq. (\ref{Lagrangian}))
enhances the quark-meson coupling and moves up 
the single quark state energy.
Thus the
contribution from a single quark state increases due to the stronger 
coupling but 
less states are accessible to be summed over up to a given energy.
Similarly, the increase of $\alpha$ suppresses the contribution of a single 
quark state since the lower Dirac components of the quark wavefunctions are 
increasing. 
But it also reduces the quark state energy, so we have more states to sum over.

The main results of this paper can be summarized as follows:
1) Strange quark polarization is a very
clean and robust prediction of the chiral potential model. Up to second order
the only contribution arises from the diagram of Fig. 1. $\Delta s$ depends
only weakly on the
model parameters, and our calculation shows further that the variation of
these parameters does not influence $\Delta s$ too much,
provided we sum over the intermediate quark state up
to a given energy.
2) The contribution from the intermediate excited quark states are important.
It is not enough to
restrict the intermediate state to the ground or the first few states.
3) Among the time-ordered diagrams, the quark-antiquark pair creation and
annihilation ``$Z$'' diagrams are significant. It is the ``$Z$'' diagrams
(Figs. 4C and 4D) that introduce a negative value for $\Delta s$ in the nucleon,
while the intermediate
negative-energy states (Fig. 4B) gives a fairly large positive contribution.
The importance of the pair
creation and annihilation contribution to $\Delta q$ has also been noticed
by some of us previously in a valence and sea quark mixing model \cite{model'}.
4) Our numerical result is quite consistent with experiments
($\Delta s$($Q^2=3$GeV$^2$)$=-0.10\pm 0.01 \pm \cdots$, where the second $\pm$
sign represents further sources of error, principally the low $x$ extrapolation
\cite{Ellis}) and lattice QCD calculations
($\Delta s=-0.12(1)$ \cite{Latt1},$ -0.109(30)$ \cite{Latt2}), and is also
consistent with a schematic calculation in the context of chiral quark
model by Cheng and Li \cite{CL}.
To the best of our knowledge this is first time that
$\Delta s$ is consistently calculated up to the one-loop level
in a quark model.

This work is supported by the CNSF (19675018), CSED, CSSTC,
the DFG (FA67/25-1), and the DAAD.

\end{multicols}

\begin{references}
\bibitem{spin} For a review of the nucleon spin problem, see 
H.Y. Cheng, Int. J. Mod. Phys. A {\bf 11}, 5109 (1996) (hep-ph/9607254); 
R.L. Jaffe, Phys. Today {\bf 48}(9), 24 (1995); 
U. Stiegler, Phys. Rep. {\bf 277}, 1 (1996)

\bibitem{Gutsche} T. Gutsche and D. Robson, Phys. Lett. B {\bf 229}, 333 (1989).

\bibitem{Thomas} See, e.g.,A.W. Thomas, Advan. Nucl. Phys. {\bf 13}, 1 (1984).

\bibitem{flux1} G. Parisi, R. Petronzio, F. Rapuano,
        Phys. Lett. B {\bf 128}, 418 (1983).

\bibitem{flux2} J.D. Stack, Phys. Rev. D {\bf 29}, 1213 (1984).

\bibitem{alpha1} M. Luscher, Nucl. Phys. {\bf B180}, 317 (1981).

\bibitem{alpha2} S. Itoh, Y. Iwasaki, T. Yoshie, 
	Phys. Rev. D {\bf 33}, 1806 (1986). 

\bibitem{Cornell} E. Eichten, K. Gottfried, T. Kinoshita, K.D. Lane, T.M. Yan,
	Phys. Rev. D {\bf 17}, 3090 (1978); 
	Phys. Rev. D {\bf 21}, 203  (1980).  

\bibitem{Bali} G.S. Bali, K. Schilling, A. Wachter,
	Phys. Rev. D {\bf 56}, 2566 (1997). 

\bibitem{Chen} X.S. Chen, X.B. Chen, A. Faessler, T. Gutsche, F. Wang,
	hep-ph/0005143.

\bibitem{model'} D. Qing, X.S. Chen, and F. Wang, 
	Phys. Rev. C {\bf 57}, R31 (1998);
	Phys. Rev. D {\bf 58}, 114032 (1998). 

\bibitem{Ellis} J. Ellis, hep-ph/9611208.

\bibitem{Latt1} S.J. Dong, J.P. Laga\"e, and K.F. Liu, 
	Phys. Rev. Lett. {\bf 75}, 2096 (1995).
	
\bibitem{Latt2} M. Fukugita, Y. Kuramashi, M. Okawa, and A. Ukawa, 
	Phys. Rev. Lett. {\bf 75}, 2092 (1995).
	
\bibitem{CL} T.P. Cheng and L.F. Li, Phys. Rev. Lett. {\bf 74}, 2872 (1995).
\end{references}
\end{document}